\documentclass{jltp}
\usepackage{graphicx} 

\title{Energy Scales in the Local Magnetic Excitation Spectrum of 
       YBa$_2$Cu$_3$O$_{6+y}$}

\author{Jan~Brinckmann$^1$ and Patrick~A.~Lee$^2$}

\address{{}$^1$Institut f\"{u}r Theorie der Kondensierten Materie,
  Universit\"{a}t Karlsruhe, \\
  D-76128 Karlsruhe, Germany \\
  {}$^2$Department of Physics, Massachusetts Institute of Technology, \\
  Cambridge, Massachusetts 02139}

\runninghead{J.~Brinckmann, P.~A.~Lee}{Local Magnetic Excitation 
  Spectrum of YBa$_2$Cu$_3$O$_{6+y}$}

\begin{document}

\maketitle

\begin{abstract} \sloppy
The wave-vector integrated dynamical spin susceptibility
$\chi_{2D}(\omega)$ of YBa$_2$Cu$_3$O$_{6+y}$ cuprates is considered.
$\chi_{2D}$ is calculated in the superconducting state from a
renormalized mean-field theory of the $t$--$t'$--$J$-model, based on
the slave-boson formulation. Besides the well-known ``41\,meV
resonance'' a second, much broader peak (`hump') appears in ${\rm
Im}\chi_{2D}$\,. It is caused by particle--hole excitations across the
maximum gap $\Delta^0$\,. In contrast to the resonance, which moves to
lower energies when the hole filling is reduced from optimal doping,
the position of this `hump' at $\approx 2\Delta^0$ stays almost
unchanged.  The results are in reasonable agreement with inelastic
neutron-scattering experiments.

\end{abstract}

\renewcommand{\textfraction}{0.0}
\renewcommand{\topfraction}{1.0}
\renewcommand{\bottomfraction}{1.0}

\section{INTRODUCTION}

The most prominent feature in the magnetic excitation spectrum of
YBa$_2$Cu$_3$O$_{6+y}$ (YBCO) and Bi$_2$Sr$_2$CaCu$_2$O$_{8+\delta}$
(BSCCO) cuprates is the so-called ``41\,meV
resonance''\cite{ros91,moo93,fon95,bou96,he01,mesot01pre} at the
antiferromagnetic (AF) wave vector ${\bf q}=(\pi,\pi)$\,. Its energy
$\omega_{res}$ is $\approx 40$\,meV in optimally doped samples and
decreases with underdoping\cite{dai96,fon97,fong00,daimook01} down to
$\omega_{res}\approx 24$\,meV\,. Recently the magnetic response has
also been studied by averaging the neutron-scattering data over the
in-plane 2D Brillouin zone\cite{fong00,daimook99}\,. The resulting
local magnetic excitation spectrum ${\rm Im}\chi_{2D}(\omega)$ shows
the above-mentioned resonance and a second, hump-like feature at an
energy $\omega_{hump}$ above $\omega_{res}$\,. In contrast to
$\omega_{res}$\,, $\omega_{hump}$ depends only weakly on the doping
level. Within the calculation to be presented in the following the
`hump' is naturally explained by particle--hole (ph) excitations
across the maximum d-wave gap $\Delta^0$\,. The energy
$\omega_{hump}\sim 2\Delta^0$ comes out almost independent of
doping. The resonance, on the other hand, emerges from a ph-bound
state in the magnetic (spin-flip) channel and shows a strong doping
dependence.

\section{MODEL AND MEAN-FIELD THEORY}

Our starting point is the doped Mott insulator. We study the
$t$--$J$-model on a simple square lattice of Cu-3d orbitals for each
of the two coupled CuO$_2$ layers (planes) in YBCO or BSCCO\,:
\begin{equation} \label{eqn-model}
  H = - \sum_{\nu,\nu',\sigma}t_{\nu\nu'}
          \widetilde{ c}^\dagger_{\nu \sigma}
          \widetilde{ c}_{\nu' \sigma} +
    \frac{1}{2}\sum_{\nu,\nu'}
    J_{\nu\nu'}\vec{S}_\nu \vec{S}_{\nu'}
    \;.
\end{equation}
In the subspace with no doubly occupied orbitals, the electron
operator on a Cu-lattice site $\nu$ is denoted 
$\widetilde{ c}_{\nu \sigma}$
with spin index $\sigma=\pm 1$\,; 
$\vec{S}_\nu$ is the spin-density operator. 
A Cu-site is specified through $\nu\equiv [i,l]$\,, where $i= 1\ldots
N_L$ indicates the Cu-position within one CuO$_2$-plane and $l=1,2$
selects the layer. $t_{\nu\nu'}$ denotes the effective intra- and
inter-layer Cu--Cu-hopping matrix elements, and $J_{\nu\nu'}$ the
antiferromagnetic super exchange .  To deal with the constraint of no
double occupancy, the standard auxiliary-particle formulation
$\widetilde{ c}_{\nu \sigma} = 
  b^\dagger_\nu f_{\nu \sigma}$
is used. The fermion $f^\dagger_{\nu \sigma}$ creates a singly
occupied site (with spin $\sigma$), the ``slave'' boson
$b^\dagger_\nu$ an empty one out of the (unphysical) vacuum
$b_\nu|0\rangle = f_{\nu \sigma}|0\rangle = 0$\,.
The constraint now takes the form
$Q_\nu = 
    b^\dagger_\nu b_\nu + \sum_\sigma 
    f^\dagger_{\nu \sigma} f_{\nu \sigma} = 1$\,.
In mean-field theory the constraint is relaxed to its thermal average
$\langle Q_\nu \rangle = 1$\,.
Together with the number $x$ of doped holes per Cu-site,
it fixes the fermion and boson densities to 
\begin{equation} \label{eqn-dens}
    (1 - x) = \sum_\sigma 
        \langle f^\dagger_{\nu \sigma}
        f_{\nu \sigma} \rangle 
    \;,\;\;\;\;
    x = 
      \langle b^\dagger_\nu b_\nu \rangle
    \;.
\end{equation}

The derivation of mean-field equations is presented in Ref.\
\onlinecite{bri01pre}\,. The dynamical spin susceptibility is given in
units of $(g\mu_B)^2$ as
\begin{displaymath}
  \chi({\bf q}, q_z, \omega)= 
    \chi^{+}({\bf q},\omega) \cos^2(\frac{d}{2}q_z) +
    \chi^{-}({\bf q},\omega) \sin^2(\frac{d}{2}q_z) 
    \;,
\end{displaymath}
where ${\bf q}$ is the in-plane wave vector, $d$ denotes the distance
of CuO$_2$ planes within a double-layer sandwich. For the even $(+)$
and odd $(-)$ mode susceptibilities a RPA-type expression is obtained,
\begin{equation}  \label{eqn-birpa}
  \chi^{\pm}({\bf q},\omega)= 
    \frac{ \chi_{p}^{irr}(\omega) }
      { 1 + \widetilde{J}^{\pm}({\bf q})\chi_{p}^{irr}(\omega) }
\end{equation}
The irreducible part $\chi^{irr}$ consists of a particle--hole (ph)
bubble of fermions as is known from BCS theory,
\begin{equation}  \label{eqn-bubb}
  \chi^{irr}_p(\omega) = 
    \frac{1}{2N_L}\sum_{{\bf k},\tilde{p}_z}
    \sum_{s, s'=\pm 1}
    \frac{1}{8}\left[ 1 + 
      ss'\frac{\varepsilon \varepsilon' + \Delta\Delta'}{E E'}
               \right]
    \frac{f(s'E') - f(s E)}{\omega + sE - s'E' + i0_+}
\end{equation}
with $p\equiv({\bf q},p_z)$ and $p_z= \{0, \pi\}$ for the modes $\{+,
-\}$\,. Boson excitations do not enter $\chi$ on mean-field
level. Fermions obey an effective dispersion
$\varepsilon\equiv\varepsilon({\bf k},\tilde{p}_z)$\,, 
$\varepsilon'\equiv\varepsilon({\bf k}+{\bf q},\tilde{p}_z + p_z)$\,, 
\begin{displaymath}
  \varepsilon({\bf k},\tilde{p}_z) =
    - 2 \widetilde{ t}[\cos(k_x) + \cos(k_y)]
    - 4 \widetilde{ t}'\cos(k_x) \cos(k_y)
    - \widetilde{ t}^\perp({\bf k}) e^{i \tilde{p}_z}
\end{displaymath}
and d-wave gap function
$\Delta\equiv\Delta({\bf k},\tilde{p}_z)$\,,
$\Delta'\equiv\Delta({\bf k}+{\bf q},\tilde{p}_z + p_z)$\,,
\begin{displaymath}
  \Delta({\bf k},\tilde{p}_z) = 
    \frac{\Delta^0}{2}[\cos(k_x) - \cos(k_y)] + 
    \Delta^{\perp 0} e^{i \tilde{p}_z}
\end{displaymath}
These enter the usual quasi-particle energy 
$E= \sqrt{\varepsilon^2 + \Delta^2}$\,,
$E'= \sqrt{{\varepsilon'}^2 + {\Delta'}^2}$\,.
Formally, vertex corrections to the simple bubble Eq.\
(\ref{eqn-bubb}) have to be taken into account. However, these have
almost no effect in the energy-range below $2\Delta^0$ and are
therefore ignored in the following\cite{bri98}\,.

From Feynman's variation principle the effective hopping parameters
are determined as\cite{bri01pre}
$\widetilde{ t}\approx x\,t + 0.15 J$\,,
$\widetilde{ t}'= x\,t'$\,,
$\widetilde{ t}^\perp({\bf k})\approx x\,t^\perp({\bf k})$\,.
For the bare nearest and next-nearest neighbor hopping we assume 
$t=2J$\,, $t'=-0.45t$\,,
and for the inter-plane hopping\cite{chasud93,okand94}
$\displaystyle  t^\perp({\bf k})= 
    2 t^\perp[ \cos(k_x) - \cos(k_y) ]^2 + t^\perp_0$
with $t^\perp= 0.1t$ and $t^\perp_0= 0$\,.  We assume an in-plane
superconducting order parameter $\Delta^0$ with equal amplitude and
phase in both layers. The self-consistent solution of the mean-field
equations then leads to a vanishing inter-plane gap
$\Delta^{\perp 0}=0$\,. 

Magnetic excitations in the superconducting phase are described by
quasi particles (the fermions) in a BCS-type d-wave pairing
state. These propagate with effective hopping parameters $\widetilde{
t}$\,, $\widetilde{ t}'$ strongly reduced from the bare parameters $t,
t'$ by the small Gutzwiller factor $x$\,. The bubble Eq.\
(\ref{eqn-bubb}) describes spin-flip ph-excitations of these
particles, which are subject to the mode-dependent final-state
interaction in Eq.\ (\ref{eqn-birpa}),
\begin{equation}  \label{eqn-effj}
  \widetilde{J}^{\pm}({\bf q})= 
    \alpha J({\bf q}) \pm J^\perp
    \;,\;
    J({\bf q})= 2J[ \cos(q_x) + \cos(q_y) ]
\end{equation}
The inter-plane exchange is chosen as $J^\perp= 0.2J$\,. The
destruction of the antiferromagnetic (AF) state of the 1/2-filled
system by hole doping\cite{leefeng88,iga92,kha93,kim99} is missing in
mean-field theory. The necessary correlations of fermions and bosons
are not contained, and the AF order vanishes at an unphysically high
doping level\cite{inaba96} $x_c^0\approx 0.22$\,. Therefore we
assume\cite{bri01pre,bri99} a renormalization $J\to \alpha J$ of the
in-plane nearest--neighbor exchange. Using $\alpha= 0.35$ reduces
$x_c^0$ down to $x_c\approx 0.03$\,, which is consistent with
experiment and makes the study of underdoped systems possible. Note
that the above-mentioned renormalization $t\to\widetilde{ t}$ of the
quasi-particles comes out of the self-consistent calculation, whereas
$J\to\alpha J$ is a phenomenological model. Our assumption of 
$\alpha$ being independent of doping leads to an AF correlation
length\cite{bri01pre}
$\xi_{AF}(x)\sim 1/\sqrt{x - x_c}$ at $T\to 0$\,,
which agrees with known experimental\cite{thu89} and
theoretical\cite{singlen92} results. 

\section{RESULTS}

\begin{figure}[h]
  \centerline{
    \includegraphics[width=0.7\hsize]{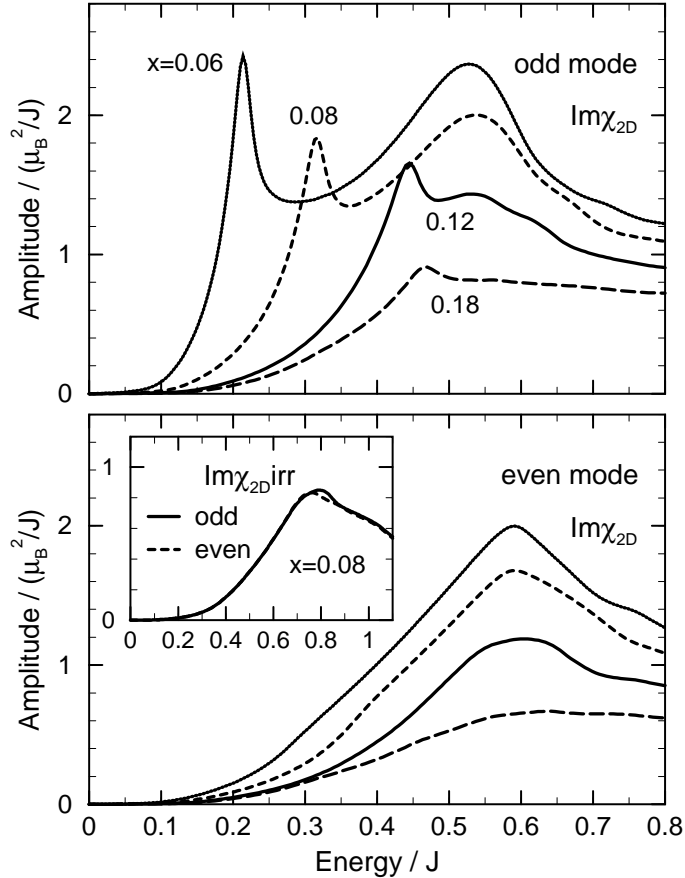}  }
  \caption[\ ]{
    Wave-vector ${\bf q}$ integrated 
    odd- and even-mode susceptibilities ${\rm Im}\chi_{2D}$
  for hole filling $x= 0.06\ldots 0.18$\,. Parameters are $t=2J$\,,
  $t'=-0.45t$\,, $t^\perp=0.1t$\,, $J^\perp=0.2J$\,. Curves are
  calculated with a damping FWHM$=0.04J\approx 5$\,meV\,. 
  {\bf Inset:} 
  ${\bf q}$-integrated bubble spectrum ${\rm Im}\chi_{2D}^{irr}$ for
  $x=0.08$\,. The maximum is located close to $2\Delta^0= 0.78J$\,. 
    }
  \label{fig-chi2d}
\end{figure}

From the susceptibility Eq.\ (\ref{eqn-birpa}) the local magnetic
excitation spectrum is determined from
\begin{equation}  \label{eqn-chi2d}
  {\rm Im}\chi_{2D}^{\pm}(\omega)=
    \int\!\!\!\!\int\limits_{-\pi}^\pi\!\!\frac{{\rm d}^2 q}{(2\pi)^2}
    \,{\rm Im}\chi^{\pm}({\bf q},\omega)
\end{equation}
Fig.\ \ref{fig-chi2d} shows ${\rm Im}\chi_{2D}$ for the odd and even
mode at $T\to0$ in the superconducting state. A resonance is clearly
visible in the odd $(-)$ mode, at an energy $\omega_{res}= 0.42J\approx
50$\,meV for $x=0.12$ near optimal doping. When $x$ is reduced
(underdoping) the resonance moves to lower energies and gains spectral
weight. The resonance appears at the same energy\cite{bri01pre} as in
${\rm Im}\chi^-({\bf q},\omega)$ for fixed wave vector ${\bf
q}=(\pi,\pi)$\,. In addition, both modes ${\rm
Im}\chi^\pm_{2D}(\omega)$ show a broad peak (`hump') at energies
$\omega^-_{hump}\approx\omega_{hump}^+$ above $\omega_{res}$\,. In
contrast to $\omega_{res}$ the hump-maxima $\omega^\pm_{hump}$ are
almost independent of doping, located somewhat below $2\Delta^0$
($2\Delta^0\approx 0.7J$ for $x=0.12$)\,.

\begin{figure}[h]
  \begin{center}
    \mbox{\includegraphics[width=0.5\hsize]{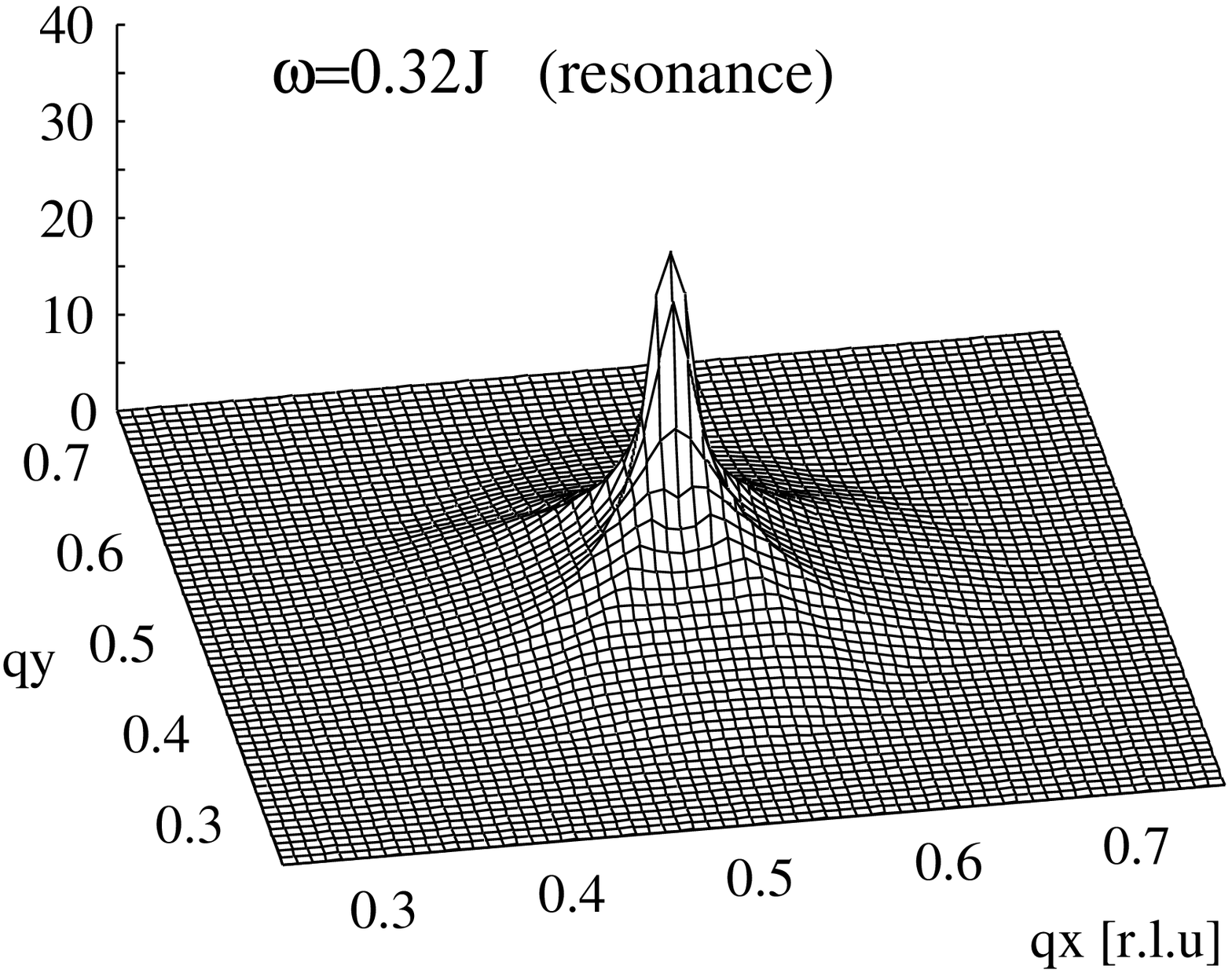}
          \includegraphics[width=0.5\hsize]{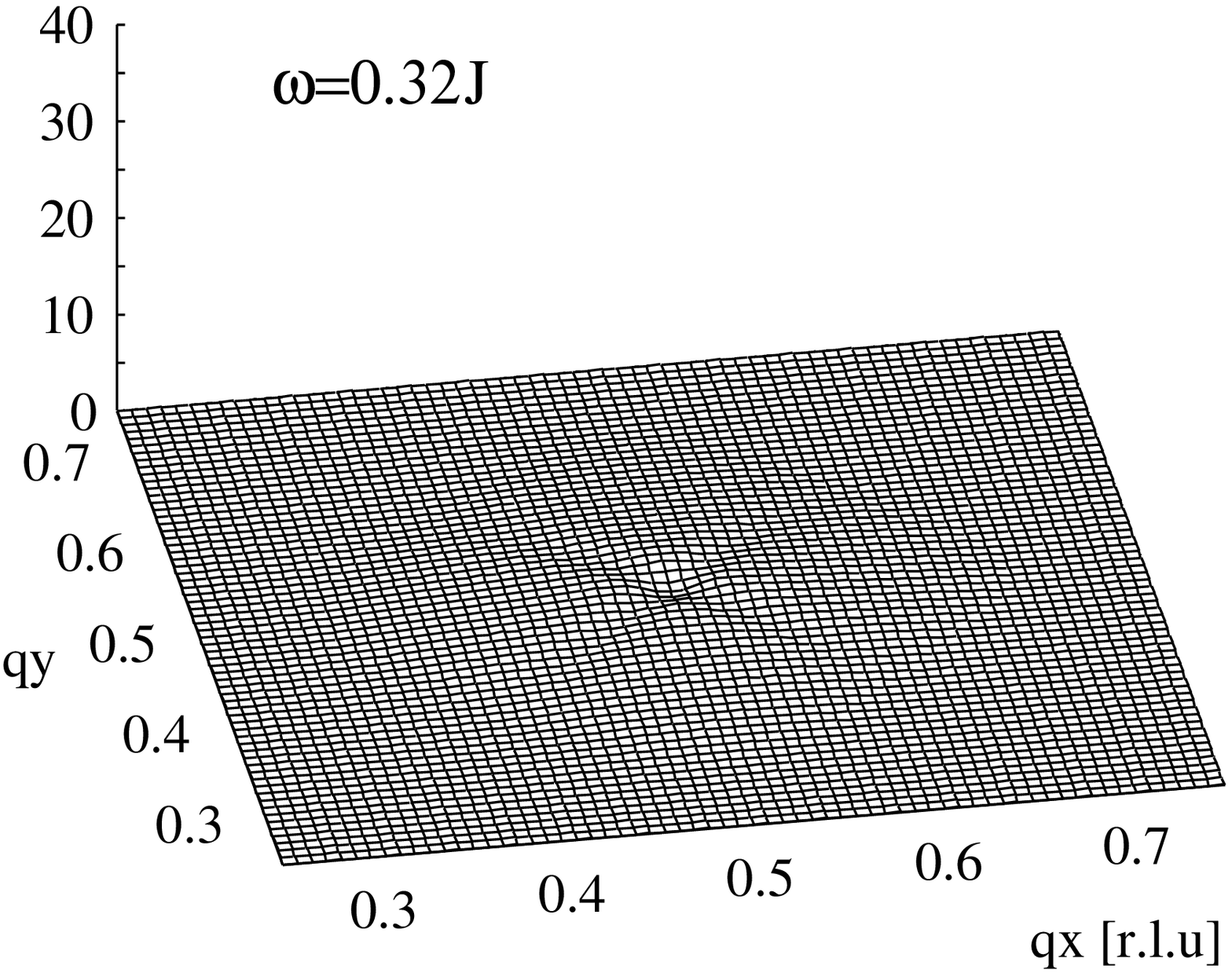}}
    \vskip 5pt
    \parbox{\hsize}{\mbox{}\hfill \fbox{odd mode} \hfill 
                    \fbox{even mode} \hfill\mbox{}}
    \vskip 5pt
    \mbox{\includegraphics[width=0.5\hsize]{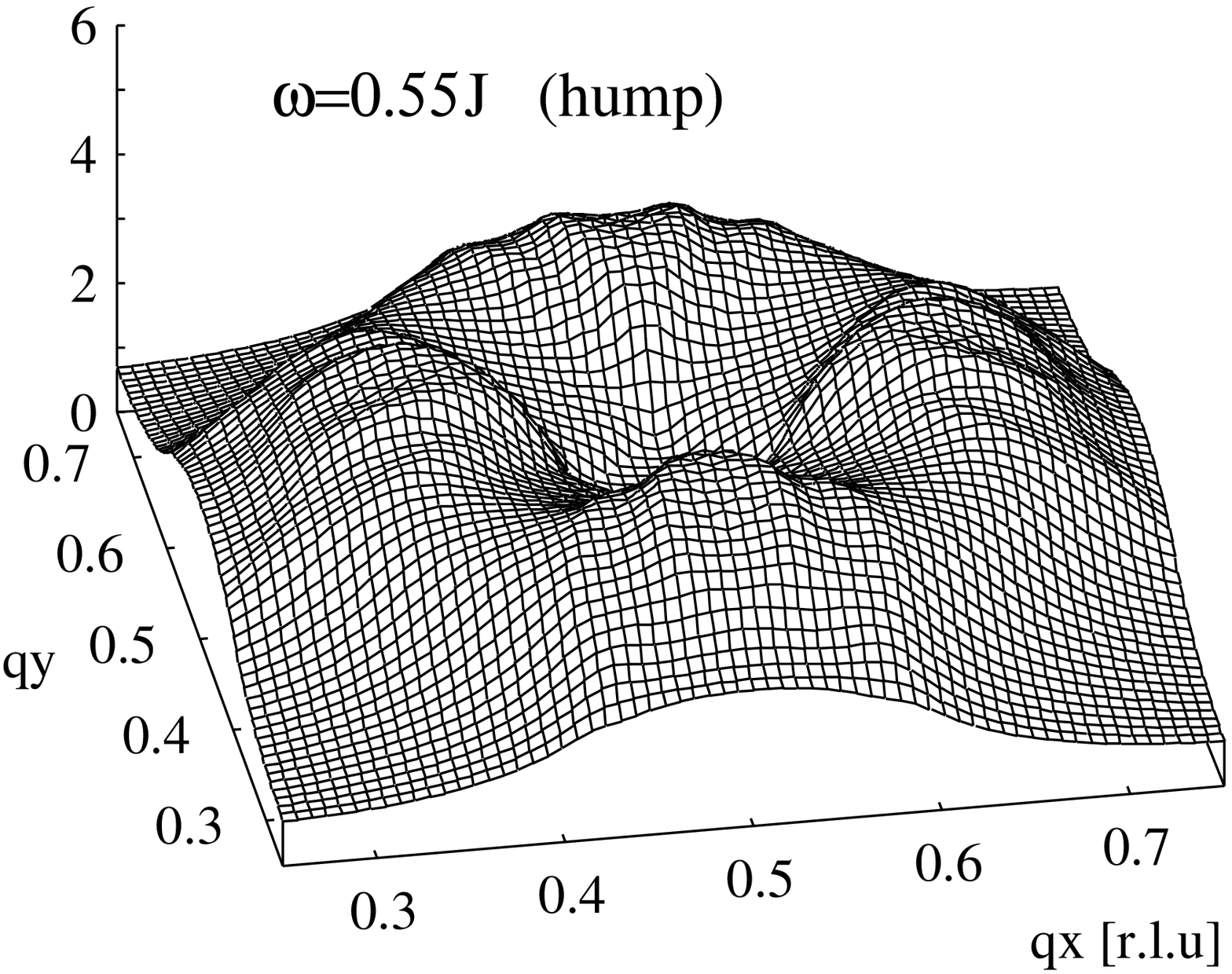}
          \includegraphics[width=0.5\hsize]{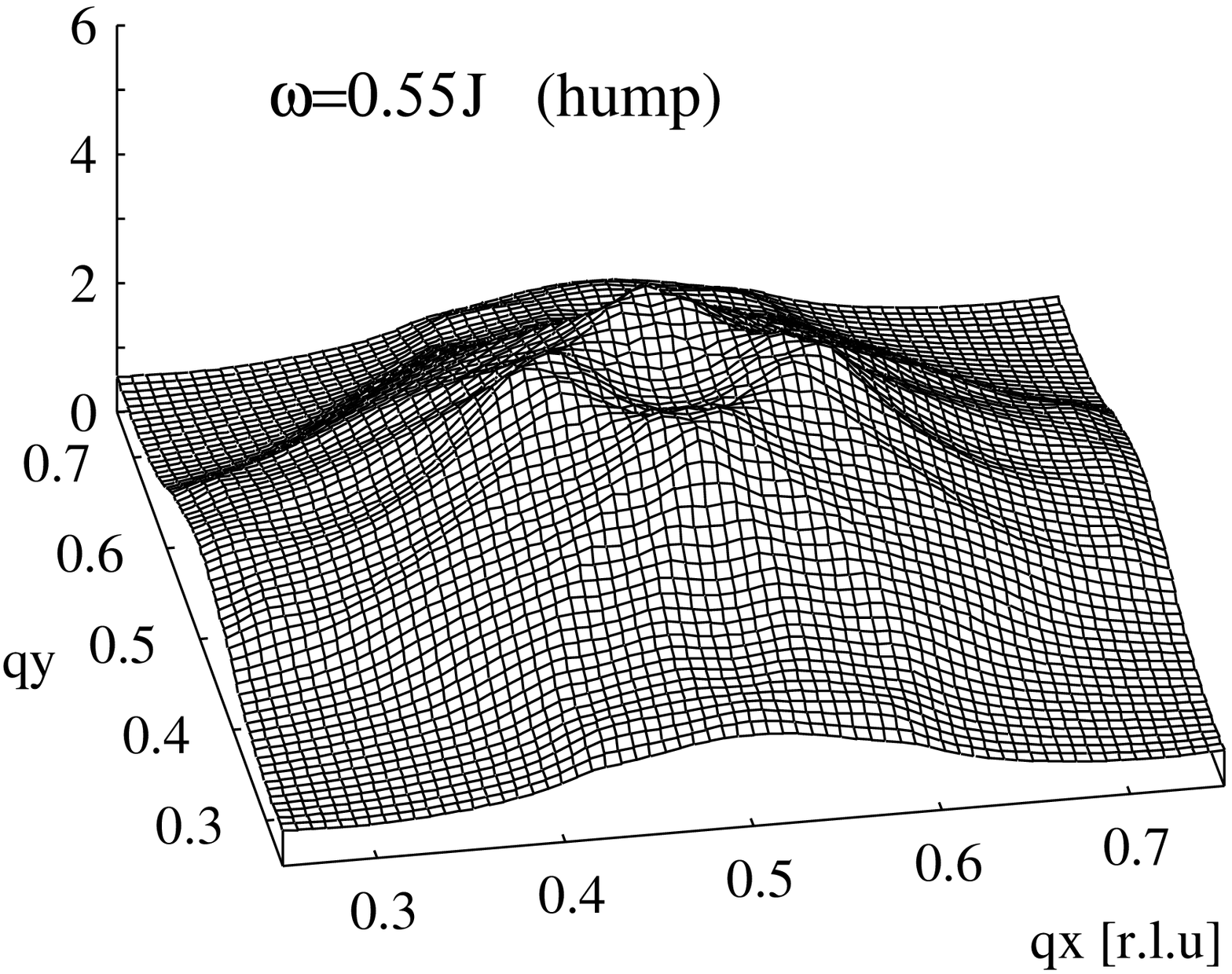}}
  \end{center}
  \caption[\ ]{
    Magnetic response ${\rm Im}\chi^{(\pm)}$ (in arbitrary units) in
  wave-vector ${\bf q}$ space for $x=0.08$\,. $q_x$\,, $q_y$ are
  measured in units of $2\pi=1$r.l.u. Parameters as in Fig.\
  \ref{fig-chi2d}\,. {\bf Top:} At the energy
  $\omega=\omega_{res}$ where the 
  resonance appears in the odd mode. {\bf Bottom:} At an energy $\omega$
  close to the `hump'-maxima in both modes. (Note the
  different amplitude scale.)
    }
  \label{fig-2dplots}
\end{figure}

The resonance emerges from a pole in Eq.\ (\ref{eqn-birpa}) at
wave-vector $(\pi,\pi)$ and energy $\omega_{res}$\,, driven by the
effective interaction Eq.\ (\ref{eqn-effj}) in the odd $(-)$
mode. Since $\omega_{res}$ is slightly below the threshold $\Omega_0$
to the particle--hole (ph) continuum, the resonance appears undamped,
i.e., as a $\delta$-function. Due to the inter-layer coupling
$J^\perp$ the interaction Eq.\ (\ref{eqn-effj}) is weaker in the even
$(+)$ mode, and the resonance in ${\rm Im}\chi^+$ is shifted up into
the ph-continuum, becoming almost suppressed. Consequently, in
wave-vector space a sharp peak around $(\pi,\pi)$ is visible only in
the odd $(-)$ mode. This is demonstrated in the top panel of Fig.\
\ref{fig-2dplots}\,. The bottom panel of that figure shows
the magnetic response at a higher energy close to the hump-maxima
$\omega^\pm_{hump}$\,. The intensity is much reduced compared to the
resonance. However, the magnetic excitations at $\approx\omega_{hump}$
occupy almost the whole 2D Brillouin zone, and despite their small
amplitude they contribute to the wave vector integrated susceptibility
Eq.\ (\ref{eqn-chi2d})\,. The `hump' can be traced back to
ph-excitations across the maximum gap $\Delta^0$\,: At ${\bf
q}=(\pi,\pi)$ the irreducible particle--hole bubble ${\rm
Im}\chi^{irr}({\bf q},\omega)$ shows a log-type van~Hove singularity
(vHs) at $\omega=2\Delta^0$\,, remnant of the density of states of the
d-wave superconductor. Moving off $(\pi,\pi)$ this vHs splits into
three vHs that disperse very weakly throughout the Brillouin zone,
leading to a soft maximum (`hump') at $2\Delta^0$ in the ${\bf
q}$-integrated (local) bubble spectrum ${\rm
Im}\chi^{irr}_{2D}(\omega)$\,. This is shown in the inset of Fig.\
\ref{fig-chi2d}\,. When the final-state interaction Eq.\
(\ref{eqn-effj}) is taken into account, the resonance is obtained in
the odd mode, and the hump is pulled down to
$\omega^-_{hump}<\omega^+_{hump}<2\Delta^0$\,. Also is the doping
dependence of $\Delta^0$ compensated: $\omega^\pm_{hump}$ are
independent of doping, while $\Delta^0$ increases with underdoping.

\section{COMPARISON TO EXPERIMENT}

Two experimental groups studied the wave-vector integrated magnetic
response ${\rm Im}\chi^\pm_{2D}$ in underdoped YBCO\,.  Refs.\
\onlinecite{daimook99,hay98} reported a line shape for YBCO$_{6.6}$\,,
which agrees quite well with our theoretical result for $x\le
0.08$\,. A `hump' in ${\rm Im}\chi^+_{2D}$ (even) appears at $\approx
100$\,meV\,, ${\rm Im}\chi^-_{2D}$ (odd) shows a similar structure at
a somewhat lower energy $\approx 90$\,meV\,. The well-known resonance
appears only in ${\rm Im}\chi^-_{2D}$ at $34$\,meV\,. In Refs.\
\onlinecite{fong00,bou97} two underdoped samples YBCO$_{6.7}$ and
YBCO$_{6.5}$ have been studied. In the even (``optical'') mode of
YBCO$_{6.7}$ a hump appears around $70$\,meV\,, whereas the odd
(``acoustic'') mode shows a weak hump-like structure at $\approx
55$\,meV\,, separated from the resonance at $33$\,meV\,. These
features tend to move up in energy in the more underdoped sample
YBCO$_{6.5}$\,, while the resonance in ${\rm Im}\chi^-_{2D}$ shifts
down to $25$\,meV\,.

Although the detailed experimental line shapes are not unique, the
qualitative features of our calculation are found in the
neutron-scattering spectra. In particular we reproduce the different
dependencies on doping level of the resonance at $\omega_{res}$ in the
odd mode and the hump-like feature at $\omega^\pm_{hump}$ in both
modes. Also is $\omega^-_{hump}$ of the odd mode lower than the
$\omega^+_{hump}$ of the even mode. Theory and experiments can also be
compared quantitatively. The measured neutron-scattering
intensities\cite{fong00,daimook99} are of the same order as the
theoretical ones in Fig.\ \ref{fig-chi2d} (using $J=120$\,meV\,, i.e.,
$1\mu_B^2/J= 8.3\mu_B^2/$eV). The maximum of the hump in the even, odd
mode in Fig.\ \ref{fig-chi2d} occurs at $\omega^{+,-}\approx 0.6J,
0.53J= 72$\,meV, $64$\,meV\,, in good agreement with the measurements
Ref.\ \onlinecite{fong00} on YBCO$_{6.7}$ at low temperature.

\begin{figure}[h]
  \centerline{
    \includegraphics[width=0.6\hsize]{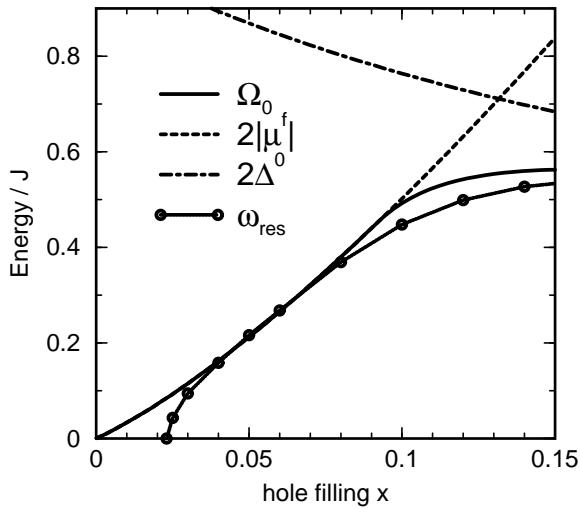}  }
  \caption[\ ]{
    $2|\mu^f|$\,, $2\Delta^0$\,, the ph-threshold $\Omega_0$\,, and
  the resonance energy $\omega_{res}$ as
  function of doping at $T\to 0$\,. 
    }
  \label{fig-dope}
\end{figure}
\section{CONCLUSION}

The local magnetic excitation spectrum ${\rm Im}\chi^\pm_{2D}(\omega)$
is characterized by two energy scales that behave differently with
hole filling $x$ (doping). The resonance energy $\omega_{res}$ follows
the particle--hole threshold $\Omega_0$\,, which in underdoped systems
is determined by the chemical potential $\mu^f$ of the fermions as
$\omega_{res}\le\Omega_0=2|\mu^f|$\,. The maxima of the humps, on the
other hand, are determined by the gap $\Delta^0$ through
$\omega^\pm_{hump}\sim 2\Delta^0$\,. When $x$ is reduced from optimal
doping, $|\mu^f|$ and therefore $\omega_{res}$ decrease quickly, while
$\Delta^0$ increases. This is displayed in Fig.\ \ref{fig-dope}\,. It
has been noted above that the mean-field theory describes magnetic
excitations in terms of quasi particles (QP) (the fermions) with a
reduced Fermi velocity $\widetilde{ v}_F\approx (x +
0.15J/t)\,v_F$\,. Hence in underdoped systems the QP's chemical
potential comes out (much) smaller than the gap, $|\mu^f|<\Delta^0$\,,
and thus determines the scale for $\omega_{res}$\,. This leads to the
observed decoupling of the resonance energy from the gap $\Delta^0$\,,
while $\Delta^0$ is still visible through the `humps' in the local
spectrum ${\rm Im}\chi^\pm_{2D}$\,.

\section*{ACKNOWLEDGMENTS}
This work has been supported by the Deutsche Forschungsgemeinschaft
through SFB\,195 and the NSF under MRSEC Program No.\
DMR\,98-08941\,. 

\sloppy


\begin{thebibliography}{10}

\bibitem{ros91}
J. Rossat-Mignod {\it et~al.}, Physica B {\bf 169},  58  (1991).

\bibitem{moo93}
H.~A. Mook {\it et~al.}, Phys.~Rev.~Lett.~ {\bf 70},  3490  (1993).

\bibitem{fon95}
H.~F. Fong {\it et~al.}, Phys.~Rev.~Lett.~ {\bf 75},  316  (1995).

\bibitem{bou96}
P. Bourges, L.~P. Regnault, L. Sidis, and C. Vettier, Phys.~Rev.~B {\bf 53},
  876  (1996).

\bibitem{he01}
H. He {\it et~al.}, Phys.~Rev.~Lett.~ {\bf 86},  1610  (2001).

\bibitem{mesot01pre}
J. Mesot {\it et~al.},   (2001), preprint cond-mat/0102339.

\bibitem{dai96}
P. Dai {\it et~al.}, Phys.~Rev.~Lett.~ {\bf 77},  5425  (1996).

\bibitem{fon97}
H.~F. Fong, B. Keimer, D.~L. Milius, and I.~A. Aksay, Phys.~Rev.~Lett.~ {\bf
  78},  713  (1997).

\bibitem{fong00}
H.~F. Fong {\it et~al.}, Phys.~Rev.~B {\bf 61},  14773  (2000).

\bibitem{daimook01}
P. Dai, H.~A. Mook, R.~D. Hunt, and F. Do{\u g}an, Phys.~Rev.~B {\bf 63},
  054525  (2001).

\bibitem{daimook99}
P. Dai {\it et~al.}, Science {\bf 284},  1344  (1999).

\bibitem{bri01pre}
J. Brinckmann and P.~A. Lee,   (2001), preprint cond-mat/0107138.

\bibitem{bri98}
J. Brinckmann and P.~A. Lee, J.~Phys.~Chem.~Solids {\bf 59},  1811  (1998).

\bibitem{chasud93}
S. Chakravarty, A. Sudb{\o}, P.~W. Anderson, and S. Strong, Science {\bf 261},
  337  (1993).

\bibitem{okand94}
O.~K. Andersen, O. Jepsen, A.~I. Lichtenstein, and I.~I. Mazin, Phys.~Rev.~B
  {\bf 49},  4145  (1994).

\bibitem{leefeng88}
T.~K. Lee and S.-P. Feng, Phys.~Rev.~B {\bf 38},  11809  (1988).

\bibitem{iga92}
J. Igarashi and P. Fulde, Phys.~Rev.~B {\bf 45},  12357  (1992).

\bibitem{kha93}
G. Khaliullin and P. Horsch, Phys.~Rev.~B {\bf 47},  463  (1993).

\bibitem{kim99}
D.~H. Kim and P.~A. Lee, Ann.~Phys.~(N.Y.) {\bf 272},  130  (1999).

\bibitem{inaba96}
M. Inaba, H. Matsukawa, M. Saitoh, and H. Fukuyama, Physica C {\bf 257},  299
  (1996).

\bibitem{bri99}
J. Brinckmann and P.~A. Lee, Phys.~Rev.~Lett.~ {\bf 82},  2915  (1999).

\bibitem{thu89}
T.~R. Thurston {\it et~al.}, Phys.~Rev.~B {\bf 40},  4585  (1989).

\bibitem{singlen92}
R.~R.~P. Singh and R.~L. Glenister, Phys.~Rev.~B {\bf 46},  11871  (1992).

\bibitem{hay98}
S.~M. Hayden {\it et~al.}, Physica B {\bf 241--243},  765  (1998).

\bibitem{bou97}
P. Bourges {\it et~al.}, Phys.~Rev.~B {\bf 56},  R11439  (1997).

\end{thebibliography}
\end{document}